# Rabi oscillations and photocurrent in quantum-dot tunnelling junctions

**Boris Fainberg** [*,1,2] **and Abraham Nitzan** [2]

[1] Faculty of Sciences, Holon Institute of Technology, Golomb street 52, 58102 Holon, Israel
[2] School of Chemistry, Tel-Aviv University, 69978 Tel-Aviv, Israel

* Corresponding author: e-mail fainberg@hit.ac.il, Phone: +972 3 502 6576, Fax: +972 3 502 6619

Motivated by the experiments by Zrenner et al. [Nature **418**, 612 (2002)], we study the influence of relaxation processes on converting Rabi oscillations in a strongly biased single-quantum-dot photodiode into deterministic photocurrents. We show that the behavior of a quantum dot with different tunnel rates for electron and holes is qualitatively different from that with the equal tunnel rates: in the latter case the current shows attenuating oscillations with the Rabi frequency. In contrast, for different electrons and holes tunnelling rates, the frequency of these oscillations diminishes, and they disappear beyond a definite asymmetry threshold. We give an analytical solution of the problem and a numerical example showing a different behaviour of the transferred charge in the small attenuation limit for equal and different tunnel rates for electrons and holes.

**1 Introduction** Quantum-dot (QD) and molecular conduction nanojunctions have been under intense study for some time [1-3]. The possible characterization and control of such systems using light has been recently discussed [4-7]. In an earlier work we have developed a theory for light-induced current by strong optical pulses in tunnelling nanojunctions [7]. We considered a molecular bridge represented by its highest occupied and lowest unoccupied states. We took into account two types of couplings between the bridge and the metal leads: electron transfer coupling that gives rise to net current in the biased junction and energy transfer interaction between excitations on the bridge and electron-hole formation in the leads. We have proposed an optical control method based on the adiabatic rapid passage for enhancing charge transfer in unbiased junctions where the bridging molecule is characterized by a strong charge-transfer transition. The method is robust, being insensitive to pulse area and the precise location of resonance that makes it suitable for a molecular bridge even for inhomogeneously broadened optical transition. In the absence of inhomogeneous broadening another procedure based on the π-pulse excitation can be applied. Recently Zrenner et al. [1] have demonstrated that Rabi oscillations between two excitonic energy levels of an InGaAs QD placed in a photodiode can be converted into deterministic photocurrents [Fig. 1]. This device can function as an optically triggered single-electron turnstile. However,

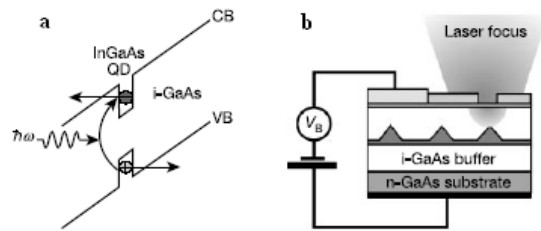

**Figure 1** a) Intrinsic region of a biased single-quantum-dot photodiode: Coherent optical π-pulse excitation applied to a QD in the ground state $|1\rangle$ leads to coherent generation of a single electron-hole pair. The photoionization of this state by tunnelling separates the electron-hole pair and results in a deterministic photocurrent. b) Schematic view of a single-QD photodiode on the basis of a GaAs n-i-Schottky diode. CB – conduction band, VB – valence band.

its efficiency for converting Rabi flopping into photocurrent is reduced at low bias voltage [10] when dephasing in the main is related to electron-phonon interaction [8,9]. In this mechanism, any π-pulse can induce a passage of one elementary charge *e* provided that the subsequent tunnelling process occurs with probability 1. Strong enough bias is needed to overcome the electron-hole attraction. At the same time the yield of photocurrent generation will be re-duced by competing processes such as radiative and non-radiative recombination of electrons and holes.

In the experiment of Ref. [1] on strongly biased single-QD photodiode the tunnelling time $\tau_{tunnel}$ ($< 10$ ps) is much shorter than low-temperature dephasing times in self-assembled InGaAs QDs ($\tau_{0,dephase} > 500$ ps). This makes photocurrent generation by Rabi oscillations relatively efficient. Below we study the influence of relaxation processes on this process at rather large bias voltages, where electron/hole tunnelling to the leads is the main relaxation mechanism.

**2 Equations of motion for the pseudospin vector and their solution** We consider a bridge characterized by two single electron orbitals $|1\rangle$ and $|2\rangle$ that are positioned below and above the equilibrium Fermi-level, respectively. The bridge interacts with the external radiation field $(1/2)E(t)\exp(-i\omega t) + c.c.$ characterized by the pulse envelope $E(t)$ and carrier frequency $\omega$. Bearing in mind the situation shown in Fig.1, Eqs. (41)-(43) and (45) of Ref. [7] can be written as

$$dn_m/dt = (-1)^m \text{Im}[\Omega^*(t)\tilde{p}_M] + \Gamma_{Mm}(\delta_{1m} - n_m) \quad (1)$$

$$d\tilde{p}_M/dt = -i\Delta\,\tilde{p}_M + i(\Omega(t)/2)(n_1 - n_2)$$
$$\quad\quad -(1/2)(\Gamma_{M1} + \Gamma_{M2})\tilde{p}_M \quad (2)$$

$$I_L = en_2\Gamma_{M2} \quad (3a)$$

$$I_R = e(1-n_1)\Gamma_{M1} \quad (3b)$$

where $n_m$ is the electron population in state $m$ ($m=1,2$), $\tilde{p}_M$ is the slowly varying amplitude of the QD polarization,

$$\Gamma_{Mm} = \frac{2\pi}{\hbar}\sum_k |V_{km}^{(M)}|^2 \delta(\varepsilon_k - \varepsilon_m) \quad (4)$$

is the corresponding tunnelling rate for the bias induced asymmetry shown in Fig.1 (we assume that hole tunnelling with rate $\Gamma_{M1}$ takes place to the right electrode and electron tunnelling with rate $\Gamma_{M2}$ is to the left electrode), $\varepsilon_j$ is the energy of state $j$ and states $\{k\}$ are single electron states of the reservoir (i-GaAs). $\Omega(t) = dE(t)/\hbar$ is the Rabi frequency, $d$ is the transition dipole moment characterizing the optical $|1\rangle \leftrightarrow |2\rangle$ transition, $\Delta = (\varepsilon_2 - \varepsilon_1)/\hbar - \omega$ is the detuning of the pulse frequency from the bridge transition frequency, $I_L$ and $I_R$ are respectively the electronic current due to the coupling of state $|2\rangle$ with the left electrode and the hole current due the coupling of state $|1\rangle$ with the right electrode. Finally, $\delta_{1m}$ is the Kronecker delta. In writing Eqs.(1)-(2) we have disregarded energy transfer terms with the rate parameter

$$B_N(\varepsilon_2 - \varepsilon_1, \mu) = \frac{2\pi}{\hbar}\sum_{k\neq k'}|V_{kk'}^{(N)}|^2 \delta(\varepsilon_k - \varepsilon_{k'} + \varepsilon_2 - \varepsilon_1)$$
$$\times f(\varepsilon_k)[1 - f(\varepsilon_{k'})] \quad (5)$$

which is associated with electron-hole excitations in the electrode. (Here $f(\varepsilon_k) = 1/\{\exp[(\varepsilon_k - \mu)/k_B T] + 1\}$ is the Fermi function and $\mu$ is the chemical potential). In doing so we have assumed that $\varepsilon_2 - \varepsilon_1$ is smaller than the band-gap of (intrinsic) i-GaAs (electrodes) as depicted in Fig.1. (Note that at sufficiently large bias, energy transfer can take place with the electron and hole created on opposite sides of the QD bridge, however, we expect that for such states the matrix elements $V_{kk'}^{(N)}$ are small, and disregard this possibility as well.)

For the following analysis it is convenient to write down Eqs. (1)-(3) in terms of the Bloch vector components [11,12] $r_1 = 2\text{Re}\,\tilde{p}_M$, $r_2 = -2\text{Im}\,\tilde{p}_M$ and $r_3 = n_2 - n_1$, and the variable $\lambda = n_1 + n_2$. This basis has the advantage of revealing the symmetry properties of the Lie group SU(2). The following analysis is made for a square pulse of duration $t_p$ and height $E_0$ starting at $t=0$. Using the unitary transformation

$$\begin{pmatrix} R_1 \\ R_2 \\ R_3 \end{pmatrix} = \begin{pmatrix} \cos 2\vartheta & 0 & -\sin 2\vartheta \\ 0 & 1 & 0 \\ \sin 2\vartheta & 0 & \cos 2\vartheta \end{pmatrix} \begin{pmatrix} r_1 \\ r_2 \\ r_3 \end{pmatrix} \quad (6)$$

where $\cos 2\vartheta = \frac{\Delta}{\sqrt{\Delta^2 + \Omega^2}}$, $\sin 2\vartheta = \frac{-\Omega}{\sqrt{\Delta^2 + \Omega^2}}$, we get the Bloch equations in the basis of dressed states. They have exact solution given by the roots of a quaternary equation corresponding to the system of differential equations (1)-(2). An interesting case is when the pulse is in resonance with the QD transition energy ($\Delta=0$). In this situation we obtain

$$\frac{dR_1}{dt} = -|\Omega|R_2 - \frac{1}{2}(\Gamma_{M2} - \Gamma_{M1})\frac{\Omega}{|\Omega|}\lambda - \frac{\Omega}{|\Omega|}\Gamma_{M1} - \frac{1}{2}(\Gamma_{M2} + \Gamma_{M1})R_1$$
$$(7)$$

$$\frac{dR_2}{dt} = |\Omega|R_1 - \frac{1}{2}(\Gamma_{M2} + \Gamma_{M1})R_2 \quad (8)$$

$$\frac{d\lambda}{dt} = \Gamma_{M1} - \frac{1}{2}(\Gamma_{M2} - \Gamma_{M1})\frac{\Omega}{|\Omega|}R_1 - \frac{1}{2}(\Gamma_{M2} + \Gamma_{M1})\lambda \quad (9)$$

$$I_L = \frac{e}{2}\Gamma_{M2}(\lambda + \frac{\Omega}{|\Omega|}R_1) \quad (10)$$

where $\lambda$ is invariant under unitary transformation (6). If we transform the Bloch vector components $R_1$, $R_2$ and $\lambda$ to new magnitudes $\tilde{R}_1, \tilde{R}_2$ and $\tilde{\lambda}$ defined by

$$(R_1, R_2, \lambda) = (\tilde{R}_1, \tilde{R}_2, \tilde{\lambda})\exp[-\frac{1}{2}(\Gamma_{M2} + \Gamma_{M1})t], \quad (11)$$

Eqs.(7)-(9) are reduced to the second order differential equation

$$\frac{d^2\tilde{R}_1}{dt^2} + [\Omega^2 - \frac{1}{4}(\Gamma_{M2} - \Gamma_{M1})^2]\tilde{R}_1 = -\frac{\Omega}{|\Omega|}\Gamma_{M1}\Gamma_{M2}\exp[\frac{1}{2}(\Gamma_{M2} + \Gamma_{M1})t] \quad (12)$$

The particular solutions of the homogeneous equation corresponding to Eq.(12) are $\exp(\pm\alpha t)$ where

$$\alpha = \frac{1}{2}\sqrt{(\Gamma_{M2}-\Gamma_{M1})^2 - 4\Omega^2} = \begin{cases} i\nu, & 4\Omega^2 > (\Gamma_{M2}-\Gamma_{M1})^2 \\ 0, & 4\Omega^2 = (\Gamma_{M2}-\Gamma_{M1})^2 \\ \gamma, & 4\Omega^2 < (\Gamma_{M2}-\Gamma_{M1})^2 \end{cases} \quad (13)$$

which defines the positive numbers $\nu$ and $\gamma$. We see that the behavior of a QD with $\Gamma_{M1} \neq \Gamma_{M2}$ is qualitatively different from that of a QD with $\Gamma_{M1} = \Gamma_{M2}$ (including the case $\Gamma_{M1} = \Gamma_{M2} = 0$ which is realized in the absence of charge transfer). When $\Gamma_{M1} = \Gamma_{M2}$, $R_1$ and the current $I$ show attenuating oscillations with Rabi frequency $\Omega$. In contrast, if $\Gamma_{M1} \neq \Gamma_{M2}$, the frequency of these oscillations diminishes, and they disappear when $4\Omega^2 \leq (\Gamma_{M2}-\Gamma_{M1})^2$. In general we expect that tunnelling rates differ between different bridge levels, $\Gamma_{M1} \neq \Gamma_{M2}$.

Solving inhomogeneous Eq. (12) with the initial conditions

$$R_1(0) = -r_3(0)(-\frac{\Omega}{|\Omega|}) = -\frac{\Omega}{|\Omega|},$$

$$R_2(0) = r_2(0) = 0, \quad \lambda(0) = 1, \quad (14)$$

and using Eqs. (10), (11) and (13), we get for the electronic current in the underdamped case $[4\Omega^2 > (\Gamma_{M2}-\Gamma_{M1})^2]$

$$I_L = \frac{e}{2}\frac{\Gamma_{M2}}{\Gamma_{M1}\Gamma_{M2}+\Omega^2}\{\frac{2\Gamma_{M1}\Omega^2}{\Gamma_{M1}+\Gamma_{M2}} + \{(\Gamma_{M2}-\Gamma_{M1})$$

$$\times [\frac{\Gamma_{M1}\Gamma_{M2}+\Omega^2}{\Gamma_{M1}+\Gamma_{M2}} - \mathrm{Re}[(i\frac{\Omega^2}{2\nu} + \frac{\Gamma_{M1}\Gamma_{M2}}{\Gamma_{M1}+\Gamma_{M2}})\exp(i\nu t)]]$$

$$+ 2\nu\,\mathrm{Re}[(-\frac{\Omega^2}{2\nu} + \frac{i\Gamma_{M1}\Gamma_{M2}}{\Gamma_{M1}+\Gamma_{M2}})\exp(i\nu t)]\}$$

$$\times \exp[-\frac{1}{2}(\Gamma_{M1}+\Gamma_{M2})t]\} \quad (15)$$

If $|\Omega|\gg\Gamma_{M1},\Gamma_{M2}$, we obtain

$$I_L = \frac{e}{2}\Gamma_{M2}\{\frac{2\Gamma_{M1}}{\Gamma_{M1}+\Gamma_{M2}} + [\frac{\Gamma_{M2}-\Gamma_{M1}}{\Gamma_{M1}+\Gamma_{M2}}$$

$$-\cos(\nu t)]\exp[-\frac{1}{2}(\Gamma_{M1}+\Gamma_{M2})t]\} \quad (16)$$

that gives

$$I_L = \frac{e}{2}\Gamma_M[1-\cos(\nu t)\exp(-\Gamma_M t)] \quad (16a)$$

for $\Gamma_{M1}=\Gamma_{M2}=\Gamma_M \equiv \frac{1}{2}(\Gamma_{M1}+\Gamma_{M2})$.

The beating phenomenon is seen to be seating on an underlying background. The corresponding expressions for the hole current $I_R$ can be obtained from Eqs.(15) and (16) by the replacement $\Gamma_{M1} \to \Gamma_{M2}$ and $\Gamma_{M2} \to \Gamma_{M1}$.

In the experiment of Ref.[1] an observable quantity is the charge transferred due to an electromagnetic pulse action $Q = \int_0^\infty I(t)dt$. Using Eqs.(1) and (3a), we obtain

$$Q_L = \int_0^{t_p} I_L(t)dt + \frac{1}{\Gamma_{M2}}I_L(t_p) \quad (17)$$

If $t_p\Gamma_M \ll 1$, the second term on the right-hand side of Eq.(17) gives the main contribution to the transferred charge. Integrating Eq.(15), we get for $|\Omega|\gg\Gamma_{M1},\Gamma_{M2}$

$$Q_L = e\{\frac{\Gamma_{M1}\Gamma_{M2}t_p}{\Gamma_{M1}+\Gamma_{M2}} + \frac{\Gamma_{M2}^2+\Gamma_{M1}^2}{(\Gamma_{M1}+\Gamma_{M2})^2} - \frac{1}{2}[\frac{(\Gamma_{M2}-\Gamma_{M1})^2}{(\Gamma_{M1}+\Gamma_{M2})^2} - \cos(\nu t_p)]$$

$$\times \exp[-\frac{1}{2}(\Gamma_{M1}+\Gamma_{M2})t_p]\} \quad (18)$$

The corresponding expression for the hole charge $Q_R = \int_0^\infty I_R(t)dt$ can be obtained from Eq.(18) by the replacement $\Gamma_{M1} \to \Gamma_{M2}$ and $\Gamma_{M2} \to \Gamma_{M1}$ like before. Since Eq.(18) is symmetric with respect to $\Gamma_{M1}$ and $\Gamma_{M2}$, we get $Q_R = Q_L \equiv Q$. Indeed, using Eqs. (1) and (3), one can obtain the following equation for the magnitude $\lambda = n_1 + n_2$: $\frac{d\lambda}{dt} = -I_L/e + I_R/e$. Integrating the last equation from $t=0$ to $t=\infty$, we get

$$e[\lambda(\infty) - \lambda(0)] = -Q_L + Q_R \quad (19)$$

Eq.(19) expresses the charge conservation. Both transferred charges $Q_L$ and $Q_R$ coincide only for $\lambda(\infty) = \lambda(0)$. This is achieved when both $\Gamma_{M2}$ and $\Gamma_{M1}$ are much larger than the pulse repetition frequency that was realized in experiment [1]. Dependences of $Q$ on $t_p$ are shown in Fig. 2. Our calculations are in good agreement with the experimental results of [1,10]. One can see that the beats contrast is somewhat better for $\Gamma_{M2} \neq \Gamma_{M1}$ than in the case $\Gamma_{M1}=\Gamma_{M2}$. This can be explained by the fact that for large $t_p$, the first term on the right-hand side of Eq. (18) gives the main contribution:

$$Q \propto \frac{e}{2}\frac{\Gamma_{M1}\Gamma_{M2}}{\Gamma_M}t_p \quad (20)$$

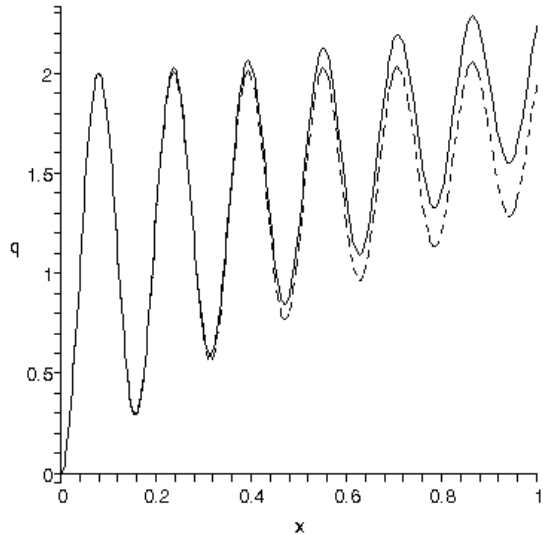

**Figure 2** The charge transferred after the completion of the pulse action as a function of its duration. q=2$Q/e$, x=$\Gamma_M t_p$, v/$\Gamma_M$ =40, $\Gamma_{M1}=\Gamma_{M2}$ (solid line) and $\Gamma_{M2}/\Gamma_{M1}$=1.9 (dashed line). The value of $\Gamma_M=(\Gamma_{M1}+\Gamma_{M2})/2$ is the same for both curves.

Indeed, for a given $\Gamma_M$ the product $\Gamma_{M1}\Gamma_{M2}$ is largest when $\Gamma_{M1}=\Gamma_{M2}$.

**3 Conclusion** In this work we have applied a theory developed by us for the light-induced current in tunnelling nanojunctions [7] to the experiments by Zrenner et al. [1] on converting Rabi oscillations in a strongly biased single-QD photodiode into deterministic photocurrents. We have shown that the behavior of a QD with different tunnel rates for electron and holes $\Gamma_{M1}$ and $\Gamma_{M2}$, respectively, is qualitatively different from that of a quantum dot with equal $\Gamma_{M1}$ and $\Gamma_{M2}$. In the latter case the current shows attenuating oscillations with the Rabi frequency. In contrast, for different tunnel rates, the frequency of these oscillations diminishes, and they disappear when $4\Omega^2 \leq (\Gamma_{M2}-\Gamma_{M1})^2$. We have obtained an analytical solution of the problem. Fig. 2 shows a somewhat different behaviour of the transferred charge in the small attenuation limit for equal and different tunnelling rates for electrons and holes.

The method considered here, and the method for enhancing charge transfer based on the adiabatic rapid passage, which has been proposed in Ref. [7], taken together enable us to realize an optically triggered single-electron turnstile based on a bridge, which is characterized by both an homogeneously or inhomogeneously broadened optical transitions.

**Acknowledgements** A.N. thanks the Israel Science Foundation, the US-Israel Binational Science Foundation and the Germany-Israel Foundation for support.